\documentclass[10pt]{elsarticle}

\usepackage{csquotes}

\usepackage[normalem]{ulem}
\usepackage{dirtytalk}
\usepackage{lineno,hyperref}
\usepackage{array, booktabs, makecell}
\usepackage[table]{xcolor} 
\usepackage{tcolorbox}
\usepackage{colortbl}
\usepackage{comment}
\usepackage{color,soul}
\usepackage{cleveref}
\usepackage{graphicx}
\usepackage{adjustbox}
\usepackage{caption}
\usepackage{subcaption}
\usepackage[title]{appendix}
\usepackage{bibentry}
\usepackage{tabularx}
\nobibliography*
\DeclareCaptionType{TextBox}
\usepackage{listings}
\usepackage{parcolumns}
\usepackage{dirtree}
\usepackage[T1]{fontenc}
\usepackage{longtable}
\usepackage{float}
\usepackage{bbding}
\usepackage{pifont}
\usepackage{wasysym}
\usepackage{amssymb}
\usepackage{rotating}
\usepackage{pstricks}
\usepackage{multirow}
\usepackage{pgfplots}
\pgfplotsset{compat=1.14}

\usepackage{tikz}
\usetikzlibrary{plotmarks}
\usepackage{verbatim}
\usepackage{xspace}
\newcolumntype{L}{>{\arraybackslash}m{13cm}}
\newcolumntype{C}[1]{>{\centering\let\newline\\arraybackslash\hspace{0pt}}m{#1}}
\newcolumntype{R}[1]{>{\raggedleft\let\newline\\arraybackslash\hspace{0pt}}m{#1}}
\modulolinenumbers[5]
\newcommand{\ie}{\textit{i.e., \xspace}}
\newcommand{\eg}{\textit{e.g., \xspace}}
\newcommand{\etal}{\textit{et al.}}

\usepackage{pgf-pie}
\usepackage{pgfplots}
\usepackage{tikz}
\definecolor{findOptimalPartition}{HTML}{D7191C}
\definecolor{storeClusterComponent}{HTML}{FDAE61}
\definecolor{dbscan}{HTML}{ABDDA4}
\definecolor{constructCluster}{HTML}{2B83BA}

\journal{Information and Software Technology}

\newcommand{\toolname}{\textsc{AntiCopyPaster}\xspace}

\newcommand{\RQone}{To what extent is the CNN model able to correctly detect the Extract Method refactoring compared to other models?}
\newcommand{\RQtwo}{Do users find \toolname and the recommended \textit{Extract Method} refactorings useful?}
\usepackage{enumitem}
\newlist{questions}{enumerate}{2}
\setlist[questions,1]{label=\bfseries RQ\arabic*.,ref=RQ\arabic*}
\setlist[questions,2]{label=(\alph*),ref=\thequestionsi(\alph*)}

\usepackage{tikz}
\usetikzlibrary{shapes.geometric, arrows}

\tikzstyle{startstop} = [rectangle, rounded corners, minimum width=3cm, minimum height=1cm,text centered, draw=black, fill=black!10]

\tikzstyle{io} = [trapezium, trapezium left angle=70, trapezium right angle=110, minimum width=3cm, minimum height=1cm, text centered,text width=2cm, draw=black]

\tikzstyle{process} = [rectangle, minimum width=3cm, minimum height=1cm, text centered, text width=3cm, draw=black]

\tikzstyle{decision} = [diamond, minimum width=3cm, minimum height=1cm, text centered,text width=2cm, draw=black]

\tikzstyle{arrow} = [thick,->,>=stealth]

\tikzstyle{database} = [cylinder, shape border rotate=90, draw=black,minimum height=2cm,minimum width=3cm, text centered,text width=0.6cm]

\usepackage{listings}

\newcommand{\lstbg}[3][0pt]{{\fboxsep#1\colorbox{#2}{\strut #3}}}
\lstdefinelanguage{diff}{
  basicstyle=\ttfamily\scriptsize,
  morecomment=[f][\lstbg{red!20}]-,
  morecomment=[f][\lstbg{green!20}]+,
  morecomment=[f][\textit]{@@},
}

\definecolor{javared}{rgb}{0.6,0,0} 
\definecolor{javagreen}{rgb}{0.25,0.5,0.35} 
\definecolor{javapurple}{rgb}{0.5,0,0.35} 
\definecolor{javadocblue}{rgb}{0.25,0.35,0.75} 
 
\usepackage{multirow} 
\usepackage{dirtytalk}
\usepackage{comment}
\usepackage{multicol}
\usepackage{varwidth} 

\definecolor{Gray}{gray}{0.80} 

\newlength\q 

        {\begin{itemize}\setlength{\itemsep}{0pt}\setlength{\parsep}{0pt}
	\setlength{\topsep}{0pt}\setlength{\parskip}{0pt}}
	{\end{itemize}}
	{\begin{enumerate}\setlength{\itemsep}{0pt}\setlength{\parsep}{0pt}
	\setlength{\topsep}{0pt}\setlength{\parskip}{0pt}}
	{\end{enumerate}}
	{\begin{enumerate}\setlength{\itemsep}{0pt}\setlength{\parsep}{0pt}
	\setlength{\topsep}{0pt}\setlength{\parskip}{0pt}\setlength{\leftmargin}{0pt}}
	{\end{enumerate}}   

\begin{document}

\begin{frontmatter}

\title{Just-in-Time Code Duplicates Extraction}
 
\author[Stevens]{Eman Abdullah AlOmar\corref{mycorrespondingauthor}\fnref{contrib}}
\cortext[mycorrespondingauthor]{Corresponding author}
\fntext[contrib]{Authors contributed equally}
\ead{ealomar@stevens.edu}

\author[HSE]{Anton Ivanov\fnref{contrib}}
\ead{apivanov_1@edu.hse.ru}

\author[JetBrains1]{Zarina Kurbatova}
\ead{zarina.kurbatova@jetbrains.com}

\author[JetBrains1]{Yaroslav Golubev}
\ead{yaroslav.golubev@jetbrains.com}

\author[RIT]{Mohamed Wiem Mkaouer}
\ead{mwmvse@rit.edu}

\author[ETS]{Ali Ouni}
\ead{ali.ouni@etsmtl.ca}

\author[JetBrains2]{Timofey Bryksin}
\ead{timofey.bryksin@jetbrains.com}

\author[RIT]{Le Nguyen}
\ead{ln8378@rit.edu}

\author[RIT]{Amit Kini}
\ead{ak3328@rit.edu}

\author[RIT]{Aditya Thakur}
\ead{at4415@rit.edu}

\address[Stevens]{Stevens Institute of Technology, Hoboken, NJ, USA}
\address[HSE]{HSE University, Moscow, Russia}
\address[JetBrains1]{JetBrains Research, Belgrade, Serbia}
\address[RIT]{Rochester Institute of Technology, Rochester, NY, USA}
\address[ETS]{ETS Montreal, University of Quebec, Montreal, QC, Canada}
\address[JetBrains2]{JetBrains Research, Limassol, Cyprus}

\begin{abstract} 

\noindent\textbf{Context:} Refactoring is a critical task in software maintenance, and is usually performed to enforce better design and coding practices, while coping with design defects. The \textit{Extract Method} refactoring is widely used for merging duplicate code fragments into a single new method. Several studies attempted to recommend \textit{Extract Method} refactoring opportunities using different techniques, including program slicing, program dependency graph analysis, change history analysis, structural similarity, and feature extraction. However, irrespective of the method, most of the existing approaches interfere with the developer's workflow: they require the developer to stop coding and analyze the suggested opportunities, and also consider all refactoring suggestions in the entire project without focusing on the development context. 

\noindent\textbf{Objective:} To increase the adoption of the \textit{Extract Method} refactoring, in this paper, we aim to investigate the effectiveness of machine learning and deep learning algorithms for its recommendation while maintaining the workflow of the developer. 

\noindent\textbf{Method:} The proposed approach relies on mining prior applied \textit{Extract Method} refactorings and extracting their features to train a deep learning classifier that detects them in the user's code. We implemented our approach as a plugin for IntelliJ IDEA called \toolname. To develop our approach, we trained and evaluated various popular models on a dataset of 18,942 code fragments from 13 Open Source Apache projects.
 
\noindent\textbf{Results:} The results show that the best model is the Convolutional Neural Network (CNN), which recommends appropriate \textit{Extract Method} refactorings with an F-measure of 0.82. We also conducted a qualitative study with 72 developers to evaluate the usefulness of the developed plugin. 

\noindent\textbf{Conclusion:} The results show that developers tend to appreciate the idea of the approach and are satisfied with various aspects of the plugin's operation.

\end{abstract}

\end{frontmatter}

\section{Introduction}

Duplicating a code fragment is the act of copying and pasting it with or without minor modifications into another section of the code base. Despite being an intuitive practice of code reuse, duplicate code brings its own challenges to software maintenance and evolution~\cite{roy2009comparison,hu2021assessing}. Recent studies have shown that duplicate code has become a problem that affects both developers and researchers. Developers can suffer from fixing a bug in a duplicate code fragment, which might then need to be applied to all of its siblings~\cite{thongtanunam2019will}. This can complicate and slow down the maintenance and lead to bug propagation. On the other hand, researchers may face problems when building machine learning models, combined with natural language processing techniques, which are designed to learn from code to support various software development practices. The threat of duplicate code can easily introduce data leakage when appearing in both training and testing data~\cite{lopes2017dejavu,allamanis2019adverse}. Therefore, removing duplicate code via refactoring has become a natural response to arising challenges~\cite{fanta1999removing}.

Refactoring is the practice of improving software quality without altering its behavior~\cite{fowler2018refactoring}. Developers intuitively refactor their code for multiple purposes: improving program comprehension, removing duplicate code, reducing complexity, dealing with technical debt, and removing code smells~\cite{Silva:2016:WWR:2950290.2950305,murphy2008breaking,ouni2016multi,mkaouer2015many}. Refactoring duplicate code consists in taking a code fragment and moving it to create a new method, while replacing all instances of that fragment with a call to this newly created method. This refactoring is known as \textit{Extract Method}.
Various studies have recommended refactorings using different motivation and base, such as improving code structure~\cite{kanemitsu2011visualization,bavota2014automating}, feature extraction~\cite{xu2017gems,yue2018automatic,aniche2020effectiveness}, reducing code duplication~\cite{yoshida2019proactive,alcocer2020improving,hotta2010duplicate,higo2008metric}, and removing the \textit{Long Method} code smell~\cite{yang2009identifying,morales2017use,tiwari2022identifying,khomh2012exploratory,palomba2014they,palomba2018diffuseness}. 

Despite their promising results, the adoption of these studies is challenged by their need to exhaustively search the entire code base to recommend proper \textit{Extract Method} refactorings. In other words, they take the entire source code of the project as input and analyze it to find any \textit{Extract Method} refactoring opportunities. However, developers may not have the privilege or the knowledge to perform a program-wide refactoring, which makes them reluctant to adopt these recommendations. Moreover, this process requires developers to find a separate time to review various refactoring suggestions that are unrelated to their current work in the code, which reduces the chances of these recommendations to be accepted in practice. Even though the \textit{Extract Method} refactoring is a built-in feature in modern programming environments, its adoption by developers is still limited even in this setting~\cite{golubev2021one}.  

\begin{figure}[t]
  \centering
  \includegraphics[width=\columnwidth]{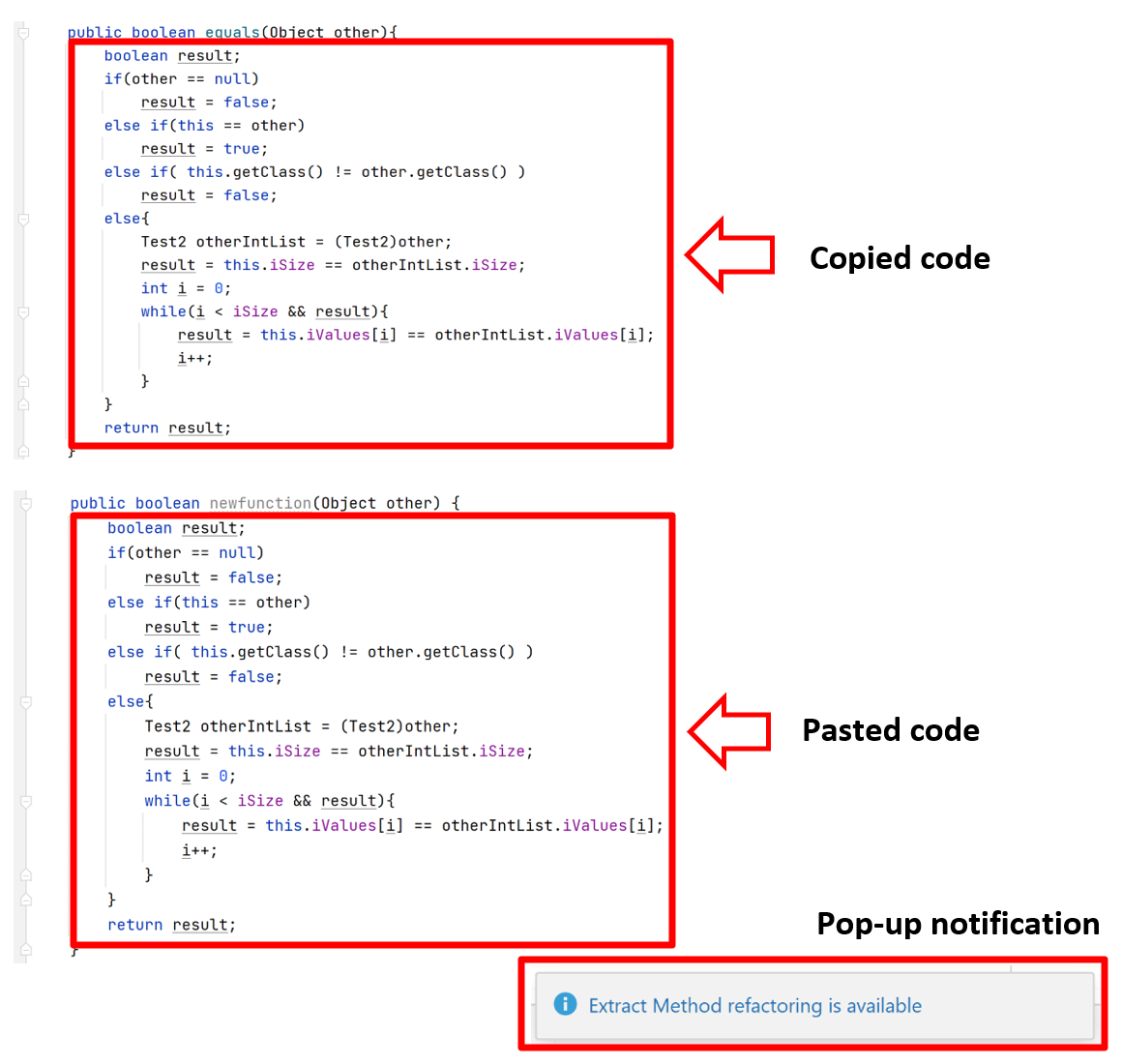}
  \caption{\textit{Extract Method} refactoring opportunity (code fragments extracted from~\cite{example}). 
  }
  \label{fig:example}
 \end{figure}

 \begin{figure*}
\centering
\begin{subfigure}{\columnwidth}
\centering\includegraphics[width=\columnwidth]{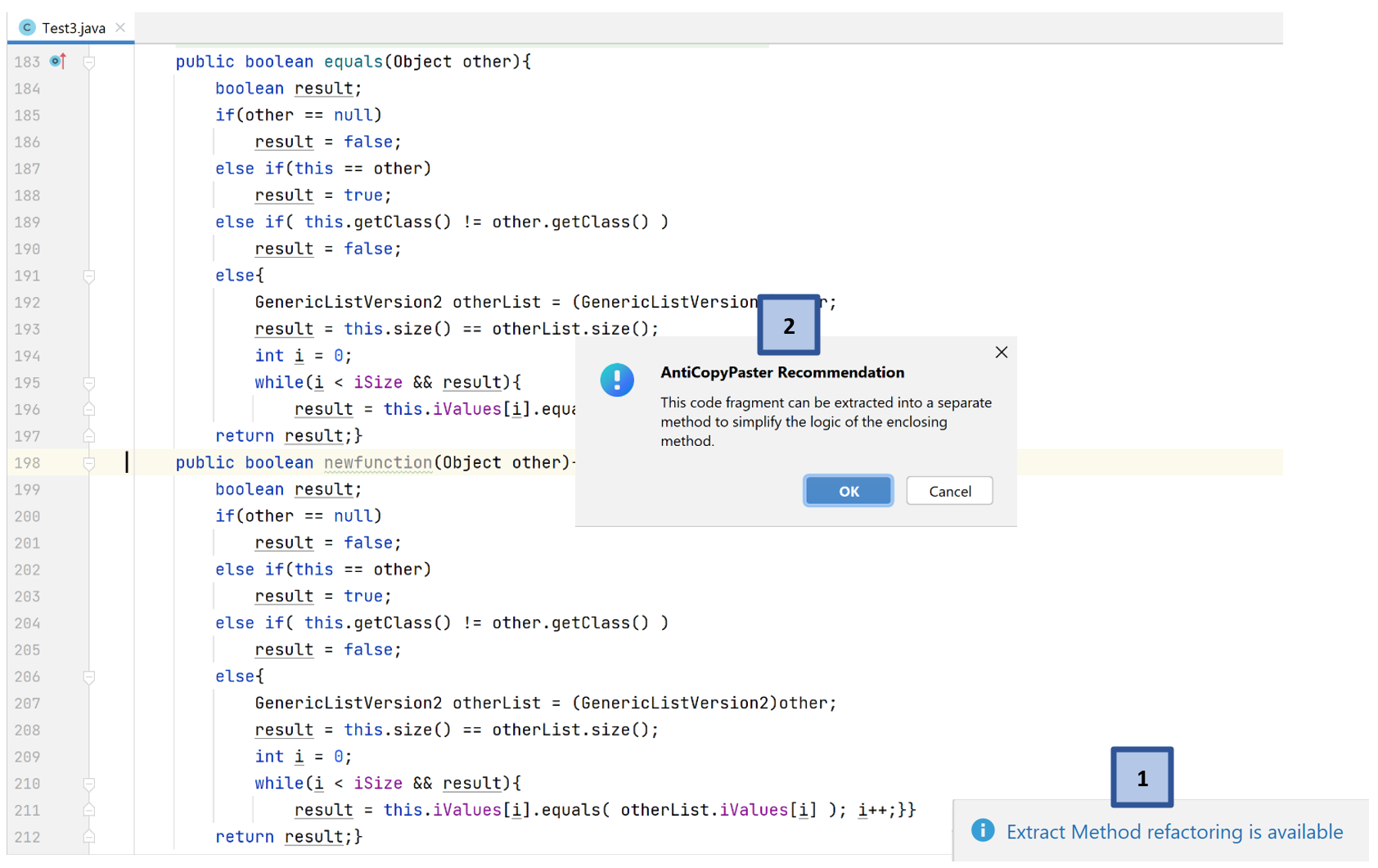}
\caption{Identification of \texttt{Duplicate Code} instances.}
\vspace{0.3cm}
\label{step1}
\end{subfigure}
\vspace{0.60cm}
\begin{subfigure}{\columnwidth}
\centering\includegraphics[width=\columnwidth]{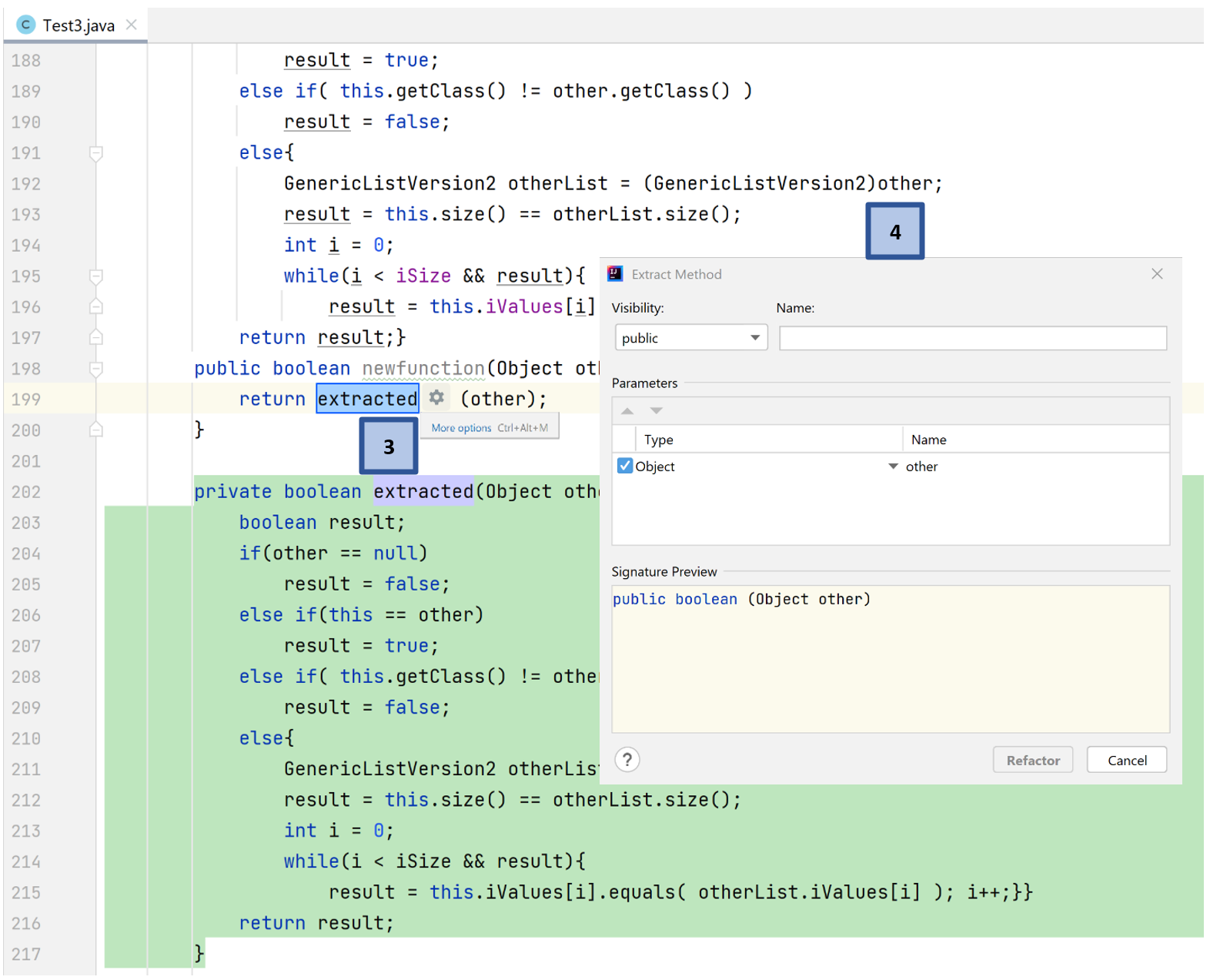}
\caption{Correction of \texttt{Duplicate Code} through the application of \textit{Extract Method} refactoring.}
\vspace{-0.5cm}
\label{step2}
\end{subfigure}
\caption{\toolname in action, showing the identified \texttt{Duplicated Code} 
 and the recommended \textit{Extract Method} refactoring.} 
\label{fig:anticopypaster-workflow}

\end{figure*}

To address the above-mentioned challenges, the goal of this paper is to aid developers with \textit{just-in-time} refactoring of duplicate code. Unlike the existing approaches that follow a posterior approach of removing accumulated duplicate code, we aim at increasing the awareness of developers \textit{while writing} their code, \ie removing duplicate code as soon as it is introduced in their code base. To do so, we design an automated approach implemented as an IntelliJ IDEA\footnote{IntelliJ IDEA: \url{https://www.jetbrains.com/idea/}} plugin called \toolname that monitors the introduction of potential duplicate code and pro-actively recommends its refactoring using the IDE's \textit{Extract Method} feature. 

As shown in Figures~\ref{fig:example} and \ref{fig:anticopypaster-workflow}, when a duplicate piece of code is pasted and is not edited for some time, a pop-up notification appears at the bottom of the screen, alerting the developer of a potential \textit{Extract Method} refactoring. The developer can choose whether to click on the notification or ignore it, until it disappears after a few seconds. If the notification is clicked, the \textit{Extract Method} feature is called with the duplicate code as input, and a refactoring preview window is opened. The developer can either apply the refactoring and suggest a name to the newly created method, or cancel the process. 

The main advantage of this solution, in comparison with previous works, is the ability to recommend \textit{just-in-time} refactorings, which increases the chances of their acceptance, since it recommends changes to a code that is (1) just edited by the developer, and (2) within the current context of development. However, not all duplicate code fragments need to be extracted, and the main challenge is being able to recommend refactoring only when the refactoring is \textit{worth it}, in order not to bother the developer by suggesting to extract random pieces of code or trivial statements.

In essence, we tackle the problem of whether the given duplicate code fragment should be extracted as a binary classification problem. First, the pasted code fragment is parsed using the IDE's Program Structure Interface (PSI)\footnote{PSI: \url{https://plugins.jetbrains.com/docs/intellij/psi-files.html}} to generate its corresponding syntactic and semantic model. This model is used to calculate a set of 78 comprehensive structural and semantic metrics, previously used in various studies recommending the \textit{Extract Method} refactoring ~\cite{aniche2020effectiveness,tiwari2022identifying,haas2016deriving,maruyama2001automated,tsantalis2011identification,sharma2012identifying,silva2014recommending,silva2015jextract,shahidi2022automated,van2021data}. Based on these metrics, a binary classification is performed to decide whether or not to suggest the refactoring. To handle the high dimensionality of the data, we design our binary classifier using a Convolutional Neural Network (CNN). 

We evaluate our approach in terms of two dimensions: \textit{correctness} and \textit{usefulness}. The first dimension evaluates the CNN's learning ability to correctly detect whether a given code fragment should be refactored. We trained and tested our model on a dataset of 18,942 code fragments mined from 13 mature Apache projects. 
Our experimental results show that our CNN model achieves high performance in the binary classification, with an F-measure of 0.82, which outperforms all other evaluated machine learning algorithms, such as Random Forest, Support Vector Machine, Naive Bayes, and Logistic Regression, while also being lightweight and convenient to use. 
  
As for the second dimension, we designed a survey that invites developers to use \toolname and reflect on its usability and usefulness. In total, 72 developers participated in the survey.
The results show that the vast majority of participants found our \toolname tool useful and were satisfied with its operation. 
Furthermore, the survey has also shown that the majority of participants are not necessarily familiar with the \textit{Extract Method} built-in IDE feature, and therefore, we foresee their usage of the plugin as an opportunity to raise the awareness of this refactoring type.

This paper extends our recently accepted tool paper~\cite{alomar2022duplicatesextractor} by providing more details about the used model and its design, discussing the used metrics and the performance of various models, and describing the full comprehensive analysis of user's feedback, including their general usage of refactorings and the open coding of their answers. To summarize, this paper makes the following key contributions:
\begin{itemize}
    \item We design and implement a novel approach called \toolname that pro-actively identifies and recommends the \textit{Extract Method} refactoring as soon as a duplicate code fragment is pasted into the file.
    \item We train, deploy, and evaluate a CNN model that has shown good performance with respect to other models, achieving an F-measure of 0.82.
    \item We provide \toolname as an open source tool publicly available for the community~\cite{Tool}. We also provide a comprehensive replication package containing the dataset, survey results, and scripts, as well as documentation on how to embed another model into the plugin, allowing researchers and practitioners to customize the tool and match their own preferences~\cite{ReplicationPackage}.
\end{itemize}

The remainder of this paper is organized as follows. Section \ref{sec:background} elaborates on the main concepts of the model. 
Section~\ref{sec:rw} reviews the existing studies related to the \textit{Extract Method} refactoring opportunities. Section~\ref{sec:approach} outlines our approach, including data collection, metrics selection, and model training. The tool implementation is discussed in Section~\ref{sec:toolimplementation}. Section~\ref{sec:results} describes the conducted evaluation and its research questions, as well as our findings. Section~\ref{Section:Threats} captures threats to the validity of our work, and we conclude the paper in Section~\ref{Section:Conclusion}.
\section{Background}
\label{sec:background}

In this section, we elaborate on the main concepts of CNNs as discussed by Kriszhevsky~\etal~\cite{krizhevsky2017imagenet}.

Convolutional neural networks (CNNs) are a type of neural network architecture that uses a series of filters to extract significant features for the purpose of classification. These filters are typically convolutional layers, mapping inputs from the previous layer to the next layer by applying trainable weights to a predefined window of data and then outputting the weighted sum of that window. Another filter that can be applied is a pooling layer. Pooling allows the accumulation of features from maps generated during the convolution. The idea of pooling is to reduce the spatial size of the representation and decrease the number of parameters and computation in the network. Typically, this is done via a max pooling layer, which again looks at a window of data from the previous input layer, and then takes the maximum value in that window to output to the next layer. Pooling layers are a computationally efficient way to distill significant features from the previous layer, since there are no weights to train. In our work, we also used traditional fully connected feedforward layers as well, which take input values, apply a trainable weight to it and feed it through an activation function before outputting.

Since neural networks may quickly overfit when trained on fewer but similar instances, there are various strategies to mitigate such overfitting. Dropout is a regularization strategy that deactivates a defined percentage of neurons in a layer at random every training epoch. Deactivating these neurons ensures that not every neuron is exposed to all the data every epoch, so it cannot overfit to superfluous patterns in the data. Also, it reduces any feature detection co-adaptations as the deactivated neurons cannot influence the retained ones.

Batch normalization is a transformation applied to the current batch of input data to the network. As the name implies, it normalizes the current input batch to take out unnecessary bias inherent in the feature data, such as some features having a different order of magnitude than others. By normalizing the input data, the CNN can train more effectively and will converge with higher accuracy.

The choice of activation function is critical to the learning rate of neural networks. It also specifies the model's type of prediction. In this paper, Rectified Linear Unit (ReLU) was selected as the activation function. The ReLU function is a piecewise function that is zero for negative values and linear for positive values. This is a classic function used in neural networks because it masks out any negatively weighted features and then gives linear importance to the positively weighted ones.

During the model's training, the loss function computes the error between the ground truth data labels and the predicted labels. This error between the actual versus the predicted labels (\ie loss) is used to update the weights of the model to maximize accuracy (minimize loss). Loss minimization is carried out by first propagating forward through the network, and applying all of the current filters and weights to the input data to calculate predictions. Then, the loss is calculated from the current predictions, before performing a backward propagation to find the gradient of each weight with respect to the loss (\ie finding the contribution of each weight to the loss). Using the gradients from each weight, the stochastic gradient descent is carried out on the loss space, which results in minimizing it.

\section{Related Work}
\label{sec:rw}

Various studies that relate to software refactoring have been of importance to both practitioners and researchers. A considerable effort has been spent by the research community on identifying and suggesting \textit{Extract Method} refactorings. The focus of these studies ranges from using program slicing techniques~\cite{maruyama2001automated,tsantalis2011identification,tsantalis2009identification} and graph representations of code~\cite{kanemitsu2011visualization,tiwari2022identifying,sharma2012identifying,shahidi2022automated} to relying on scoring functions to find the most appropriate refactoring candidates~\cite{yang2009identifying,haas2016deriving,silva2014recommending,silva2015jextract} and using machine learning techniques~\cite{xu2017gems,yue2018automatic,aniche2020effectiveness,van2021data}. We summarize the key studies in Table~\ref{Table:Related_Work_in_Extract_Method_Refatoring}.

\begin{table*}
  \centering
	 \caption{Related work in identifying the \textit{Extract Method} refactoring opportunities.}
	 \label{Table:Related_Work_in_Extract_Method_Refatoring}
  \vspace{0.5cm}
\begin{sideways}
\begin{adjustbox}{width=1.05\textwidth,center}
\begin{tabular}{lclllll}\hline
\toprule
\bfseries Study & \bfseries Year  & \bfseries Approach & \bfseries Technique  & \bfseries Tool & \bfseries Design Defect &  \bfseries Plugin? \\
\midrule
Maruyama~\cite{maruyama2001automated} & 2001 & Rule-based & Code slicing & Not mentioned & Not mentioned & No \\
Murphy-Hill \& Black~\cite{murphy2008breaking} & 2008 & Rule-based & Assertion & Not mentioned & Not mentioned & No \\ 
Tsantalis \& Chatzigeorgiou~\cite{tsantalis2009identification,tsantalis2011identification} & 2009 \& 2011 &   Rule-based & Code slicing & JDeodorant & Long Method & Yes \\
Yang~\etal~\cite{yang2009identifying} & 2009 &   Score-based & Fragment identification & AutoMeD & Long Method & No \\
Kanemitsu~\etal~\cite{kanemitsu2011visualization} & 2011 &  Graph-based & Program dependency graph &  ReAF & Not mentioned & No \\
Sharma~\cite{sharma2012identifying} & 2012 & Graph-based & Data \&  structure dependency & N/A & Not mentioned & No  \\
Silva~\etal~\cite{silva2014recommending,silva2015jextract} & 2014 \& 2015 &   Score-based & Structural similarity &  JExtract & Not mentioned & Yes \\
Charalampidou~\etal~\cite{charalampidou2016identifying} & 2016 &   Rule-based & Functional relevance & SEMI &  Long Method & No\\
Haas \& Hummel~\cite{haas2016deriving} & 2016 &  Score-based & Control \& data flow
graph  & ConQAT & Long Method & No \\
Xu~\etal~\cite{xu2017gems} & 2017 &  ML-based & Feature extraction & GEMS & Not mentioned & No \\
Yue~\etal~\cite{yue2018automatic} & 2018 &  ML-based & Feature extraction  & CREC & Code Clone & No \\
Yoshida~\etal~\cite{yoshida2019proactive} & 2019 & Rule-based & Code modification analysis & Not mentioned &  Code Clone & Yes \\
Aniche~\etal~\cite{aniche2020effectiveness} & 2020 &  ML-based &  Feature extraction &  N/A & Not mentioned & No \\
Van~der~Leij~\etal~\cite{van2021data} & 2021 &   ML-based & Feature extraction  & N/A & Not mentioned & No \\
Shahidi~\etal~\cite{shahidi2022automated} & 2022 &   Graph-based & Dependency graph analysis  & N/A & Long Method & No \\
Tiwari \&  Joshi~\cite{tiwari2022identifying} & 2022 & Graph-based & Segmentation  & N/A & Long Method & Yes\\ 
\textbf{This work} & & \textbf{DL-based} & \textbf{Feature extraction} & \textbf{\toolname} & \textbf{Duplicate Code} & \textbf{Yes} \\
\bottomrule
\end{tabular}
\end{adjustbox}
\end{sideways}
\vspace{-.3cm}
\end{table*}

Maruyama~\cite{maruyama2001automated} developed a semi-automated approach for suggesting refactorings, which decomposed the control flow graph
using block-based program slicing. This approach was later adapted and implemented by Tsantalis and Chatzigeorgiou~\cite{tsantalis2009identification} in the \textsc{JDeodorant} tool that identified \textit{Extract Method} refactoring opportunities using code slicing along with a set of rules to ensure behavior preservation after slice extraction. In a follow-up work, Tsantalis and Chatzigeorgiou~\cite{tsantalis2011identification} proposed a set of additional behavior preservation rules that exclude refactoring opportunities related to slices, the extraction of which could possibly cause a change in the program behavior. In another study, Murphy-Hill and Black~\cite{murphy2008breaking} presented three features to improve the adoption and usage of the \textit{Extract Method} refactoring, namely, selection assist, box view, and refactoring annotation. Their formative study shows that user satisfaction was significantly increased with these features. Sharma~\cite{sharma2012identifying} proposed \textit{Extract Method} candidates based on the data and the structure dependency graph. Their suggestions were obtained by eliminating the longest dependency edge in the graph.

Kanemitsu~\etal~\cite{kanemitsu2011visualization} presented a visualization method for identifying \textit{Extract Method} refactorings and introduced an implementation of the proposed method called \textsc{ReAF}. Another approach was proposed by Yang~\etal~\cite{yang2009identifying}, the authors identified fragments to be extracted from long methods. Their approach is implemented as a prototype called \textsc{AutoMeD}. The evaluation results suggested that the approach may reduce the refactoring cost by 40\%.

Silva~\etal~\cite{silva2014recommending} used a similarity-based approach to recommend automated \textit{Extract Method} refactoring opportunities that hide structural dependencies that
are rarely used by the remaining statements in the original
method. Their evaluation on a sample of 81 \textit{Extract
Method} opportunities achieved the precision and recall rates close to 50\% when detecting refactoring instances. In another study, Silva~\etal~\cite{silva2015jextract} extended their work by designing an Eclipse plugin called \textsc{JExtract} that automatically identified, ranked, and applied refactorings upon request. Inspired by the study of Silva~\etal~\cite{silva2015jextract}, Haas and Hummel~\cite{haas2016deriving} developed a scoring function aimed to decrease the complexity of code by considering the code's length and nesting depth. The evaluation against 10 experienced developers showed that they accepted 86\% of the suggested refactorings. Another study by Charalampidou~\etal~\cite{charalampidou2016identifying} shows the application of functional relevance to detect the \textit{Long Method} code smell. The authors developed a tool called \textsc{SEMI} to automate the application of this approach for Java classes.

Xu~\etal~\cite{xu2017gems} proposed a plugin called \textsc{GEMS} that used both structural and functional features of code fragments from the real-world \textit{Extract Method} refactorings to train a model to suggest refactorings. \textsc{GEMS} utilized the same extraction algorithm as \textsc{JExtract} \cite{silva2015jextract} to create code fragments for the extraction. Yue~\etal~\cite{yue2018automatic} presented a tool called \textsc{CRec} that combined static analysis and the analysis of the history of code to suggest code clones that should be extracted into a separate method. Another experiment was conducted using a pro-active clone recommendation system. Yoshida~\etal~\cite{yoshida2019proactive} designed an Eclipse plugin that tracks user code modification and constructed a system for supporting clone refactoring. When the system detects an \textit{Extract Method} refactoring being performed, it automatically searches for clones of the extracted fragment and suggests to extract them as well. However, this still requires a developer to think about whether a code clone is worth refactoring. 

Aniche~\etal~\cite{aniche2020effectiveness} used a machine learning approach that involves predicting refactorings using code, process, and ownership metrics. The resulting
models predict 20 different refactorings at the levels of a class, method, and variable with an accuracy often higher than 90\%. Another experiment that predicts refactorings was conducted using quality metrics. Van~der~Leij~\etal~\cite{van2021data} explored the recommendation of the \textit{Extract Method} refactoring at ING. They observed that machine learning models can recommend \textit{Extract Method} refactorings with high accuracy, and the user study reveals that ING experts tend to agree with most of the recommendations of the model.

More recently, Shahidi~\etal~\cite{shahidi2022automated} automatically identified and refactored the \textit{Long Method} code smells in Java code using advanced graph analysis techniques. Their proposed approach was evaluated on five different Java projects. The findings reveal the applicability of the proposed method in establishing the single responsibility principle with a 21\% improvement. In another study, Tiwari and Joshi~\cite{tiwari2022identifying} introduced \textit{Segmentation} as a graph-based technique to identify \textit{Extract Method} refactoring with the aim of achieving higher performance with fewer refactoring suggestions. The authors compared their approach against two state-of-the-art approaches, \ie \textsc{JExtract} and \textsc{SEMI}, and showed an improvement over both of them.

Overall, recommending \textit{Extract Method} refactoring opportunities has been extensively studied~\cite{aniche2020effectiveness,yoshida2019proactive,tsantalis2011identification,van2021data}. Although some of the proposed techniques utilized various code metrics as a new way for recommending \textit{Extract Method} refactoring, to the best of our knowledge, no prior studies have proposed a just-in-time automated \textit{Extract Method} refactoring tool for the purpose of specifically eliminating code duplication while using code metrics as features to predict whether a piece of code should be extracted. The \textit{just-in-time} automatic aspect is crucial since it allows developers to remain in the context of the suggestion and thus make a timely decision. This alleviates the burden of reviewing a long list of refactoring opportunities, located in code fragments that are irrelevant to development context. 

To gain a more in-depth understanding of the issue, increase the awareness of duplicate code, and increase the adoption of \textit{Extract Method} refactorings, in this paper, we develop a tool called \toolname that is fast, conformable to use, and integrated into the developer's IDE editor. Our study complements the existing efforts that are carried out to recommend \textit{Extract Method} refactorings in general~\cite{murphy2008breaking,aniche2020effectiveness,sharma2012identifying,silva2014recommending,van2021data} or specifically in order to eliminate certain code smells~\cite{yoshida2019proactive,tiwari2022identifying,haas2016deriving,tsantalis2011identification,shahidi2022automated,charalampidou2016identifying}.

\section{Approach}
\label{sec:approach}

\begin{sidewaysfigure}
  \centering
  \includegraphics[width=\textwidth]{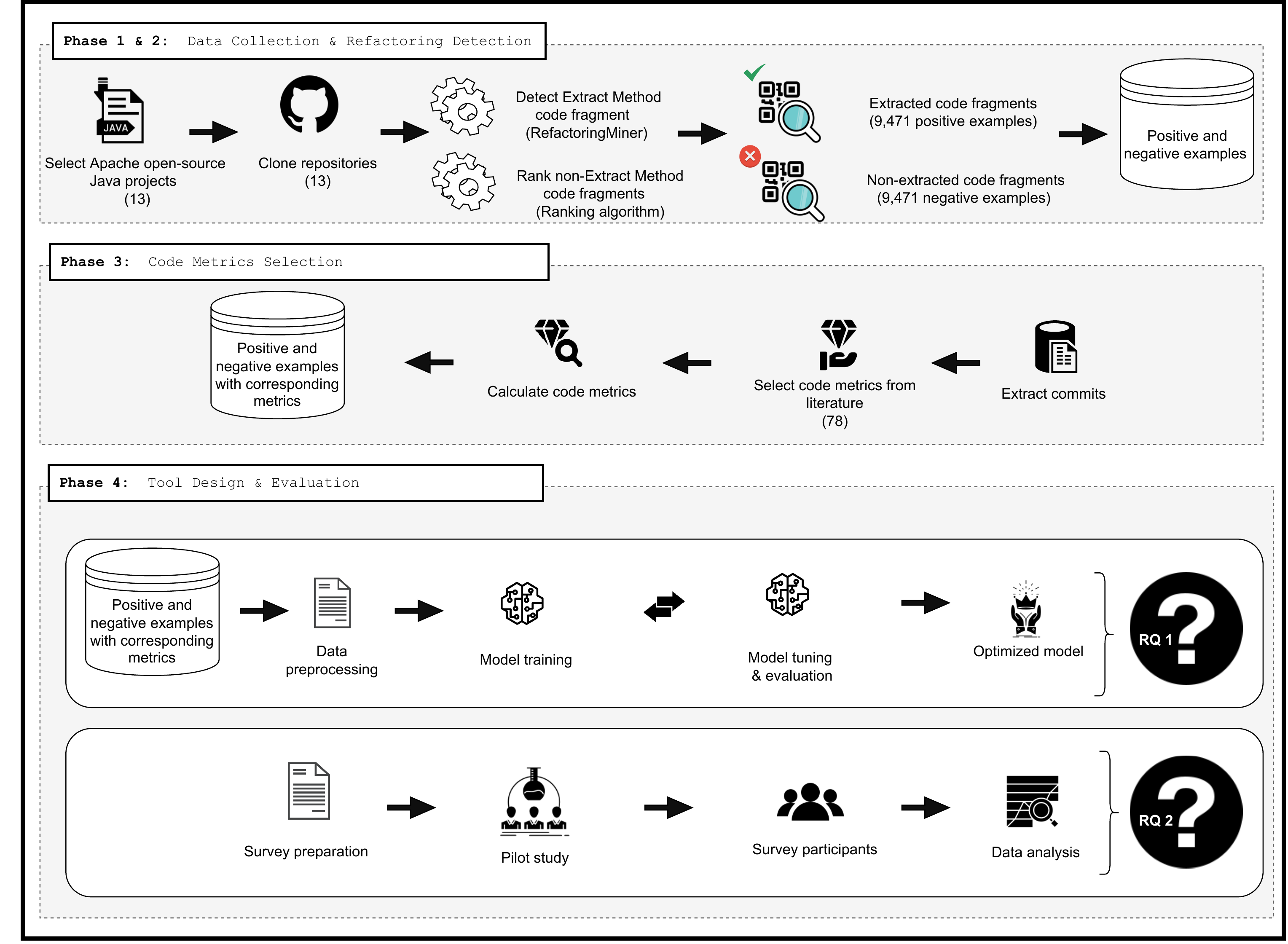}
  \caption{The overall pipeline of our work.} 
  \label{fig:pipeline}
\end{sidewaysfigure}

In a nutshell, the goal of our work is to automatically provide \textit{just-in-time} recommendations of \textit{Extract Method} refactoring opportunities as soon as duplicate code is introduced in the opened file in the IDE. Our approach takes code metrics as input and makes a binary decision on whether the code fragment has to be extracted. The present work can be divided into four phases as shown in Figure~\ref{fig:pipeline}. It consists of: (1) data collection, (2) refactoring detection, (3) code metrics selection, and (4) tool design and evaluation. The dataset, tool, and scripts utilized in this study are available in the replication package  \cite{ReplicationPackage} for extension and replication purposes. 

\subsection{Data Collection}

\begin{table}[t]
\begin{center}
\caption{The overview of the data.} 
\label{Table:DATA_Overview}
\begin{tabular}{lllll}\hline
\toprule
\bfseries Item & \bfseries Count \\
\midrule
Number of projects & 13 \\
Software quality metrics & 78 \\
Extracted code fragments (Positive Examples) & 9,471 \\
Non-Extracted code fragments (Negative Examples) & 1,000,000 \\
Selected non-Extracted code fragments (Negative Examples) & 9,471 \\
Final dataset & 18,942 \\
\bottomrule
\end{tabular}
\end{center}
\end{table}

Our first step consists of selecting 13 mature projects from the Apache Software Foundation,\footnote{Apache projects on GitHub: \url{https://github.com/apache}} which are popular open-source Java projects hosted on GitHub \cite{bavota2015apache,di2020relationship}. These curated projects were selected with respect to both project size and activity, while verifying that they were Java-based, the only language supported by RefactoringMiner \cite{tsantalis2018accurate,tsantalis2020refactoringminer}. An overview of the extracted data is provided in Table~\ref{Table:DATA_Overview}.

\subsection{Refactoring Detection}

To extract the entire refactoring history of each project, we used RefactoringMiner v2.0,\footnote{RefactoringMiner: \url{https://github.com/tsantalis/RefactoringMiner}} a widely-used refactoring detection tool introduced by Tsantalis~\etal~\cite{tsantalis2018accurate,tsantalis2020refactoringminer}. We decided to use RefactoringMiner as it has shown good results in detecting refactorings compared to other available tools (a precision of 99.8\% and a recall of 95.8\%) and is suitable for a study that requires a high degree of automation since it can be used through its external API. 

We identify methods that underwent an \textit{Extract Method} refactoring (\ie positive examples) using RefactoringMiner. In total, the tool mined 9,471 cases of \textit{Extract Method} refactorings. Specifically, we discovered \textit{Extract Method} refactorings, then traversed the history to the previous commit and took the code fragment that had been extracted. This allowed us to detect fragments that are \textit{worth} of being extracted, since they were extracted in mature projects. These refactorings are not necessarily only applied in the context of duplicate code, and thus our model learns from various contexts (\eg splitting long methods). Our model is not intended to identify duplicate code, since this is handled by another algorithm in the tool, but to evaluate whether the given duplicate code is worth refactoring. 

As mentioned in Section~\ref{sec:rw}, various approaches rank candidate code fragments for method extraction. These techniques can be also used to discover the opposite: code fragments that are \textit{less likely} to be extracted, \ie negative samples for our model. In our work, we use the ranking formula inspired by the work of Haas and Hummel~\cite{haas2016deriving}, since the authors corroborated its validity by providing a human evaluation of the results. To collect the negative samples, we start with selecting all sequences of statements that are eligible to be extracted. Then, they are ranked according to a special scoring formula proposed by Haas and Hummel~\cite{haas2016deriving} that optimizes independent code metrics. From their formula, we used three terms: \textit{statement length}, \textit{nesting depth}, and \textit{nesting area}. After ranking the fragments, according to the original work, the fragments that are more likely to be extracted will be located at the top of the list. In order to gather the ones that are less likely to be extracted, we skip the first 5\% of fragments and select the bottom 95\% of the list. We carried out this process for all 13 projects, then, to match the number of positive examples, we sampled 9,471 negative examples to constitute the final dataset.

\subsection{Code Metrics Selection}

After collecting positive and negative examples, we characterize them through various metrics. The goal of selecting metrics is to identify patterns in their values to allow distinguishing between the two classes of fragments. To do so, we gathered all the metrics that have been extensively used in previous studies~\cite{yue2018automatic,aniche2020effectiveness,haas2016deriving,caulo2020taxonomy,mkaouer2014recommendation,d2012evaluating} and then removed all the redundant metrics to avoid generating features with similar values. In machine learning, duplicated values is a well-known problem that can have adverse effects on models and cause training algorithms to overfit or inflate classification metrics~\cite{allamanis2019adverse}. In total, we selected 78 metrics that can be divided into three main categories:

\begin{enumerate}
    \item \textbf{Metrics that relate to the code fragment}: \eg length of the code fragment in symbols, \texttt{if} keyword count, etc.
    \item \textbf{Metrics that relate to the enclosing method}: \eg length of the enclosing method in lines, etc.
    \item \textbf{Coupling metrics}: \eg number of references to fields from the enclosing class in the code fragment, etc.
\end{enumerate}

The list of metrics is available in our replication package~\cite{ReplicationPackage}.

\subsection{Model Training}

\subsubsection{Dataset} To prepare the dataset, we label code fragments that underwent an \textit{Extract Method} refactoring with \say{1}, and code fragments that are less likely to be extracted with \say{0}. Our feature vectors consist of the collected code metrics values, calculated for the positive and negative examples. In total, we annotated 9,471 code fragments as positive and 9,471 code fragments as negative.

\subsubsection{CNN Binary Classification} 

We define the detection of an \textit{Extract Method} opportunity as a binary classification problem. Our intended model takes a set of metrics as input, and uses them as features to learn patterns in their values that distinguish between duplicate code fragments that are more likely and less likely to be extracted. Since the input corresponds to 78 metrics, we chose to rely on CNNs for building our model. CNNs have been proven to achieve higher performance when classifying with a high number of features \cite{gu2018recent,liu2021reproducibility}. 

\begin{figure}
  \centering
  \includegraphics[width=1.0\columnwidth]{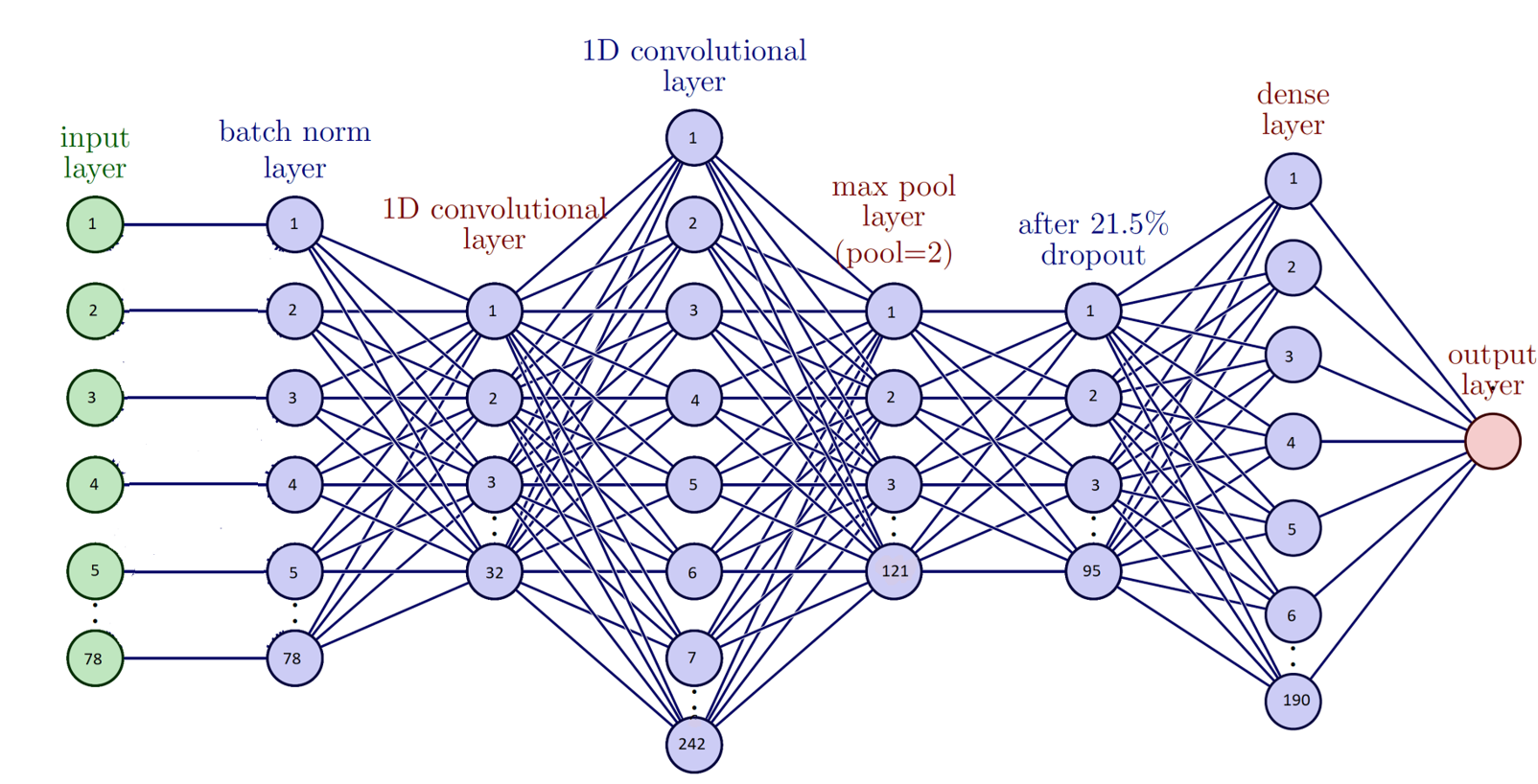}
  \caption{The architecture of the proposed CNN model.}
  \label{fig:cnn-overview}
 \end{figure}

\begin{figure}
  \centering
  \includegraphics[width=0.8\columnwidth]{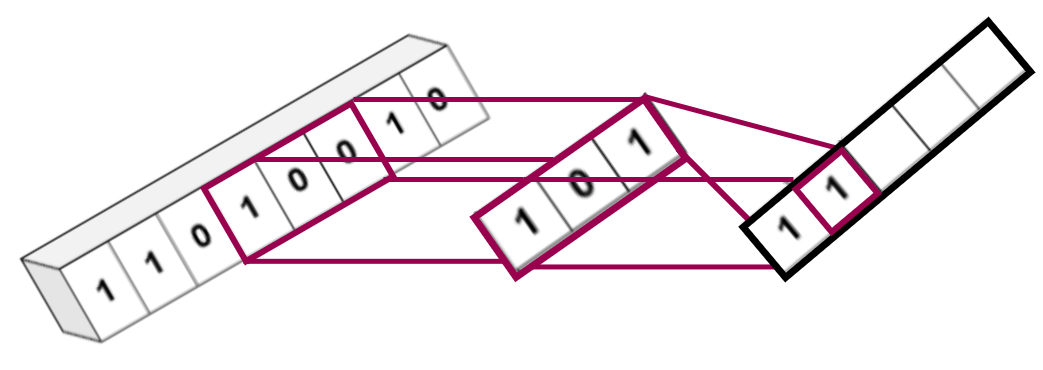}
  \caption{1D convolution with a kernel size 3 and stride 3.}
  \label{fig:cnn-window}
 \end{figure}

Our CNN model consists of multiple layers of fully connected nodes, structured into a convolutional, deconvolutional, dense layers, and a dropout stage to prevent overfitting. The visualization of this architecture can be found in Figure~\ref{fig:cnn-overview}. The input to the CNN is a vector of 78 metrics values that are batch-normalized to stabilize their distributions, through introducing additional network layers to control their mean and variance. The batch normalised inputs are fed to a convolution that reduces the feature space from 78 to 32. This convolution allows the model to adjust the weighting of the features, so the more significant ones are signal-boosted while less significant ones get suppressed (without being entirely dismissed). To do so, it takes a subset of the input vector, and applies a given weight to each element before summing them up and evaluating them using an activation function. We use the Rectified Linear Unit (ReLU) as the activation function for the convolutional layers. As an example of visualization, in Figure \ref{fig:cnn-window}, a convolutional layer with a $3\times1$ filter multiplies the filter values by the learned weights and sums them up to distill dimensionality down to the most important features. During the training, the gradients were computed using a back-propagation algorithm.

Then, the convoluted data is fed to the deconvolutional layer. This is essentially the opposite of a convolutional layer where we take one value and multiply it by the layer’s weights to increase the dimensionality of our feature space. In this layer, the most important features are boosted with higher weights once the model is trained. Thus, the training of the model adjusts the weights to boost important features and suppresses the noise by reducing the feature space to what is most significant, then immediately expanding it again to make those significant features more prominent in the network. To further refine the weighting of our set of features, we use a max-pooling layer.

The max pool layer contains a filter of size 2 and takes the largest number in the said filter. This cuts our feature space in half and only considers the most prominent values. It is also computationally efficient since there are no weights associated with this max function. Since the training data is randomized, there is a possibility of the same class instances being consecutively fed during the training, which potentially biases the model into overpredicting that class. To avoid such overtraining bias, we added a dropout section to hide a subset of the nodes every epoch.

The final layer in the CNN is a dense one, in which each node receives the input from all nodes of the previous layer. Essentially, it maps the input to the corresponding output once the weights are adjusted properly (learned). The dense layer will output to a single node, making a probabilistic decision about whether a code fragment has to be refactored, given its original inputs (metric values). The CNN has been trained to minimize the binary cross-entropy loss function that calculates the distance between the model’s predicted label and the expected one.

\subsubsection{Model Tuning} 

The purpose of this stage in the model construction process is to obtain the optimal set of classifier parameters that provide the minimized loss value; in other words, the objective of this task is to tune the hyperparameters. For our model, we optimized the batch size, number of nodes for the deconvolutional 1D layer, the dropout percentage, and the size of the dense layer. A randomized search was used to optimize these parameters \cite{bergstra2012random}. For each set of parameters, 3 epochs were run to calculate the loss value. For batch size, we generated random values between 10 to 256. For sizes of layers, we generated random numbers between 16 and 256. For the dropout percentage, we tried random percentages between 0 and 50. Lastly, for the dense layer size, we tried random sizes between 5 and 256. The size of the convolutional layer was kept constant at 32. The parameters which had the minimal loss value after 3 epochs were considered as the best set of parameters (see Table \ref{Table:Parameters}). As a result, the hyperparameter tuning has set the optimal batch size to 20, the size of the deconvolutional 1D layer to 242, the dropout percentage to 21.5, and the size of the dense layer to 190.

\begin{table}[t]
\centering
\caption{Optimal parameter values for CNN.}
\label{Table:Parameters}
\begin{tabular}{@{}ll@{}}
\toprule
\textbf{Parameter} & \textbf{Value} \\ 
\midrule
 n\_epochs & 3 \\
 batch size & 20 \\
 convolutional layer & 32 \\
 deconvolutional 1D layer & 242 \\
 dropout & 21.5 \\
 dense layer  & 190 \\ 
 \bottomrule
\end{tabular}
\end{table}

\section{Tool implementation}
\label{sec:toolimplementation}

In this section, we describe the specific implementation of our proposed approach and the trained model. \toolname is a plugin for IntelliJ IDEA, one of the most popular IDEs for Java. The plugin is composed of four main components.

\textbf{Duplicate Detector}. To detect duplicates, we use bag-of-words token-based clone detection~\cite{sajnani2016sourcerercc}. This code similarity-based approach takes a given code fragment as input, then parses all methods inside the same file, so that each method is represented as tokens. The next step is to compute the similarity between the code fragment and methods via their abstracted token representation. This approach can detect an exact match, \ie when the code fragment is a substring of the method body. The bag-of-tokens similarity also takes into account minor changes in the pasted fragment, such as reordering the sequence of code, or renaming an identifier.

Since it is possible that a code fragment will be significantly edited soon after it is pasted, in order to avoid the immediate flagging of the pasted code as duplicate, and potentially interfering with the developer's flow, we implement a \textit{delay} and place the pasted code fragment in a queue. Then, two sanity checks are executed: we check whether the pasted fragment is Java code and whether it constitutes a correct syntactic statement. To do that, the plugin tries to build a PSI tree of the fragment. A PSI (Program Structure Interface) tree is a concrete syntax tree that is used in the IntelliJ Platform to represent the code~\cite{kurbatova2021intellij}. If a PSI tree can be built and represents a valid statement, and if the duplicates still remain after the delay, the code fragment is passed to the \textit{Code Analyzer}.

\textbf{Code Analyzer.} This component takes the duplicate fragment as input and uses its PSI representation to calculate the 78 metrics that we discussed in Section~\ref{sec:approach}. The code fragment, with its corresponding vector of metrics, consitute the input to the \textit{Method Extractor}.

\textbf{Method Extractor.} This component takes as input the vector of metrics, and feeds it to the pre-trained model in order to make the binary decision of whether this code fragment is similar to the ones that have been previously refactored in the training dataset. If the binary classifier confirms the refactoring, then \textit{Refactoring Launcher} is called.

\textbf{Refactoring Launcher.} This component starts with checking if the pasted code fragment could be extracted into a separate method without any compilation errors. If all checks pass, a notification is then enabled to appear in the bottom right corner of the editor, informing the developer that an \textit{Extract Method} refactoring is recommended (see Figure~\ref{fig:example}). If the user responds to the tip, \textit{Refactoring Launcher} passes the duplicate fragment as an input to the IDE's built-in \textit{Extract Method} API, and initiates the preview window. The user previews the code change and has the choice to either confirm the refactoring, while renaming the newly extracted method, or cancel the entire process.
\section{Evaluation and Discussion}
\label{sec:results}
This section describes our empirical study aimed to evaluate the proposed approach, as well as its main results. We formulated two research questions:

\begin{questions}
    \item \RQone
    \item \RQtwo
\end{questions}

\subsection{RQ1: Correctness}

\subsubsection{Approach} To address RQ1, we explore the ability of our CNN to accurately detect \textit{Extract Method} refactoring opportunities. Furthermore, we compare the performance of our CNN model with four machine learning classifiers: Random Forest (RF), Support Vector Machine (SVM), Naive Bayes (NB), and Logistic Regression (LR).  
We selected these ML classifiers since their performance was competitive in similar binary classification problems~\cite{aniche2020effectiveness,van2021data,alomar2022documentation,alomar2021we,alomar2021toward,Levin:2017:BAC:3127005.3127016}.
To evaluate the performance of the algorithms, we use out-of-sample bootstrap validation since this validation technique yields the best balance between the bias and variance in comparison to single-repetition holdout validation~\cite{tantithamthavorn2016empirical}.

\begin{table}[h]
\begin{center}
\caption{The performance of different classifiers. Highest values are highlighted in \textbf{bold}.}
\label{Table:binary_classifiers}
\begin{adjustbox}{width=.9\columnwidth,center}
\begin{tabular}{lcccc}\hline
\toprule
\bfseries Classifier & \bfseries Precision & \bfseries Recall & \bfseries F-measure & \bfseries PR-AUC  \\
\midrule
Random Forest & \textbf{0.83} & 0.78 & 0.81 & \textbf{0.86}\\
Support Vector Machine & 0.78 & 0.74 & 0.76 & \textbf{0.86}\\
Naive Bayes & 0.72 & 0.46 & 0.56 & 0.72\\
Logistic Regression & 0.73 & 0.70 & 0.71& 0.79\\
\textbf{Convolutional Neural Network} & 0.82 & \textbf{0.82} & \textbf{0.82} & \textbf{0.86}
 \\
\bottomrule
\end{tabular}
\end{adjustbox}
\end{center}

\end{table}

\subsubsection{Results} The comparison between the classification algorithms is reported in Table \ref{Table:binary_classifiers}. 
Based on our findings, the F-measure of CNN is 82\%, higher than its competitors RF, SVM, NB, and LR, achieving 81\%, 76\%, 56\%, and 71\%, respectively. We conjecture that a proper conveyance of the semantics behind the source code would have required complex feature engineering using neural network classification strategy rather than traditional machine algorithms. This observation has been also supported by previous studies that utilized deep learning to source code analysis~\cite{zampetti2020automatically,tufano2019empirical}. 

\begin{figure}
  \centering
  \includegraphics[width=1.0\columnwidth]{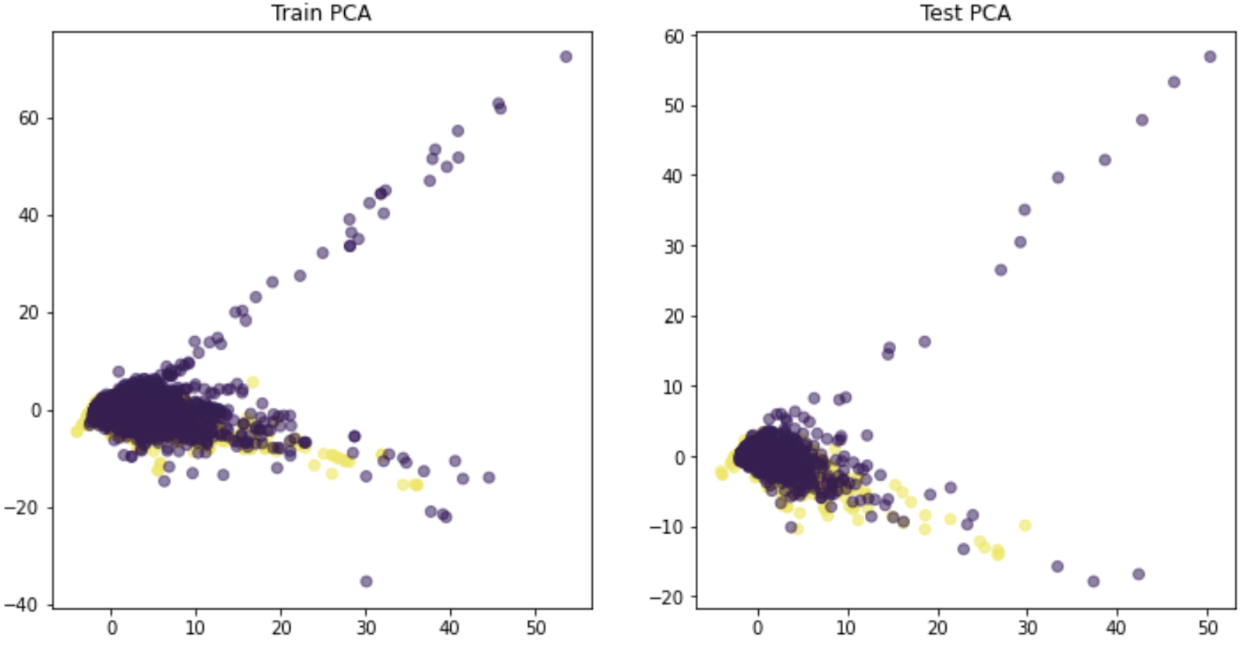}
  \caption{PCA plots between the positive (yellow) and negative (purple) classes for both training and testing data.}
  \label{fig:pca}
 \end{figure}

To further assess the efficiency of the identification of \textit{Extract Method} refactoring, we used Isolation Forest (iForest) \cite{liu2008isolation} to perform a one-class classification (anomaly detection) to evaluate the model's ability to characterize the positive examples compared to noise. iForest detects the abnormal data in sample data by looking at how many branches are needed to classify a data point and judges whether the sample is isolated based on this branching. 

To evaluate the effectiveness of iForest in \textit{Extract Method} identification, we compare iForest with other models considered in Table \ref{Table:binary_classifiers}. As can be seen, traditional classification methods obtain better performance and the Isolation Forest does not do better than random guessing (Accuracy = 0.49). Further, to better visualize the overlap of data points from the positive and negative examples, we performed the clustering on the data. Figure \ref{fig:pca} illustrates the PCA plots for both training and testing data. We can see there is high overlap between the positive (yellow) and negative (purple) classes. This indicates that the one-class classification (anomaly detection) is insufficient to classify this data because our noise overlaps too much with the positive class. Note that we used PCA for dimensionality reduction so we can get a better visualization of the clustering, but K-means clustering was carried out in the entire 78-dimensional space as well and found the same overlap (it could not distinguish between the groups better than random guessing).

Since there is no model that outperforms all the others in both precision and recall, the choice of the model can become the decision of the practitioner who is adopting the tool. For this problem, the lack of precision indicates potential recommendations of code that is not necessarily worth refactoring (\eg less complex), which would result in creating one extra method. At the same time, the lack of recall indicates missing opportunities of recommending code that is worth refactoring. We opted to deploy CNN because it not only provides the highest recall, but also delivers the best trade-off in terms of F-measure.

\begin{table}[t]
\begin{center}
\caption{Statistical comparison between different classification algorithms (McNemar’s test and Odds Ratio). `*' captures the smallest OR among 10 statistical tests.}
\label{Table:McNemar_test}
\begin{tabular}{lccc}\hline
\toprule
\bfseries Comparision & \bfseries \textit{p}-value  & \bfseries OR \\
\midrule
\textbf{CNN vs RF} &  < 0.005 &  1.45*\\
\textbf{CNN vs SVM} & < 0.005  & 2.19  \\
\textbf{CNN vs NB} &  < 0.005  & 3.59  \\
\textbf{CNN vs LR} &  < 0.005  & 2.55  \\ \hline
RF vs LR &  < 0.005 & 1.95* \\
RF vs SVM & < 0.005 & 1.54*\\
RF vs NB & < 0.005 & 2.96* \\
SVM vs NB & < 0.005 & 2.59 \\
SVM vs LR & < 0.005 & 1.77 \\
LR vs NB & < 0.005 & 2.11 \\
\bottomrule
\end{tabular}
\end{center}

\end{table}

In order to statistically compare the performance of the classification algorithms, we use the McNemar test~\cite{dietterich1998approximate} and the Odds Ratio (OR) effect size, where an OR greater than 1 indicates that the first technique outperforms the second one. We compared the performance of each pair of the classifiers by running statistical tests 10 times. Since multiple comparisons are performed, we adjusted the \textit{p}-values using the Bonferroni correction \cite{dalgaard2002analysis}. In this context, we define the following null hypothesis \(H_0\) for each test: \textit{there is no statistically significant difference between the performance of two algorithms} and our alternate hypothesis \(H_1\)  is: \textit{there is a statistically significant difference between the performance of the two algorithms}. Table \ref{Table:McNemar_test} describes the \textit{p}-values and OR obtained for each test. We tested the hypotheses at a 5\% significance level and used an adjusted alpha value (\ie 0.005) for the comparisons. As shown in the table, the McNemar test results show that null hypothesis is rejected as there are statistically significant differences (\textit{p}-value < 0.05/10) in the performance of the algorithms, with the CNN having 2.19, 3.59, and 2.55 more chances to correctly recommend \textit{Extract Method} refactoring opportunities than SVM, NB, and LR, respectively, and at least 1.45 more chances than RF. The table also reveals that the performance of the other two classifiers are statistically significant with OR greater than 1.

\begin{tcolorbox}
\textit{\textbf{Summary:} CNN outperforms traditional machine learning algorithms, having at least 1.45 more chances to recommend proper \textit{Extract Method} refactoring opportunities.} 
\end{tcolorbox}

\subsection{RQ2: Usefulness}

\subsubsection{Approach} To assess the usefulness of \toolname, we performed an external validation by involving 109 participants from the Rochester Institute of Technology, Stevens Institute of Technology, and ETS Montreal. All participants volunteered to participate in the experiment. Of the set of invited participants, 72 developers accepted  and participated in the survey (yielding a response rate of 66.1\%, which is considered high for software engineering research \cite{smith2013improving}), and 50 out of 72 participants confirmed that they executed the plugin. Table~\ref{Table:ParticipantExperiance} summarizes the developers’ experience. 88.9\% of the participants have more than 1 year of coding experience. Also, 50\% of the participants have between 1 to more than 10 years of development in either industry or open source. 

\begin{table}[b]

\begin{center}
\caption{Professional development experience of the participants in years.}
\label{Table:ParticipantExperiance}

\begin{tabular}{lllll}
\toprule
\multicolumn{1}{p{2cm}}{\textbf{Years of Experience}} &
\multicolumn{1}{p{2.8cm}}{\textbf{Professional Experience (\%)}} & 
\multicolumn{1}{p{3cm}}{\textbf{Programming Experience (\%)}} & 
\\
         \midrule 
\textbf{< 1} & 36 (50\%) & 8 (11.11\%)\\
\textbf{1-3}  & 26 (36.12\%) & 24 (33.33\%) \\ 
\textbf{4-6} & 5 (6.94\%)  & 25 (34.73\%)\\
\textbf{7-10} & 0 (0\%) & 9  (12.5\%)\\
\textbf{10+} & 5 (6.94\%) & 6 (8.33\%) \\
        \bottomrule
\end{tabular} 

\end{center}
\vspace{-.4cm}
\end{table}

As suggested by Kitchenham and Pfleeger \cite{kitchenham2008personal}, we constructed the survey to use a 5-point ordered response scale (`Likert scale') questions, 7 open-ended questions, and 8 multiple choice questions with an optional `Other' category, allowing the respondents to share thoughts not mentioned in the list. The survey consisted of 21 questions. The first part of the survey includes questions about the demographics of participants. Next, we asked about the usefulness, usability, and functionality of the proposed idea and the plugin. We provide participants with a video demonstrating how to use the plugin, along with a link to download it. 

Furthermore, we asked the participants to potentially  share with us the log of the plugin events. This log tracks the usage of the tool. Such information provides us with more detailed information about the plugin's usage over time. The log does not track any personal information related to the user or the source code. The log features the following events: 

\begin{itemize}
    \item \textbf{copyCount.} Action of copying a code fragment.
    \item \textbf{pasteCount.} Action of pasting a code fragment.
    \item \textbf{notificationCount.} Appearance of a notification about a potential refactoring opportunity.
    \item \textbf{extractMethodAppliedCount.} Acceptance of the recommendation by clicking on the notification and applying the refactoring.
    \item \textbf{extractMethodCanceledCount.} Cancelation, when the notification was clicked, but the process was canceled.
\end{itemize}

To share the log with us, the participants neede to upload an XML file, which is auto-generated by the plugin, at the end of the survey.

We analyzed the responses to open-ended questions to create a comprehensive high-level list of themes by adopting a thematic analysis approach based on guidelines provided by Cruzes~\etal~\cite{cruzes2011recommended}. Thematic analysis is one of the most used methods in Software Engineering literature~\cite{Silva:2016:WWR:2950290.2950305,alomar2022code}. This is a technique for identifying and recording patterns (or “themes”) within a collection of descriptive labels, which we call “codes”. For each response, we proceeded with the analysis using the following steps: \textit{i}) Initial reading of the survey responses; \textit{ii}) Generating initial codes (\ie labels) for each response; \textit{iii}) Translating codes into themes, sub-themes, and higher-order themes; \textit{iv}) Reviewing the themes to find opportunities for merging; \textit{v}) Defining and naming the final themes, and creating a model of higher-order themes and their underlying evidence. The above-mentioned steps were performed independently by two authors. One author performed the labeling of responses to open-ended questions independently from the other author, who was responsible for reviewing the currently drafted themes. Then, the authors met and refined the themes. It is important to note that the approach is not a single-step process. As the codes were analyzed, some of the first cycle codes were subsumed by other codes, relabeled, or dropped altogether. As the two authors progressed in the translation to themes, there was some refinement and reclassification
of data into different or new codes.

\subsubsection{Results} 
\begin{figure}
\centering 
\includegraphics[width=1.0\columnwidth]{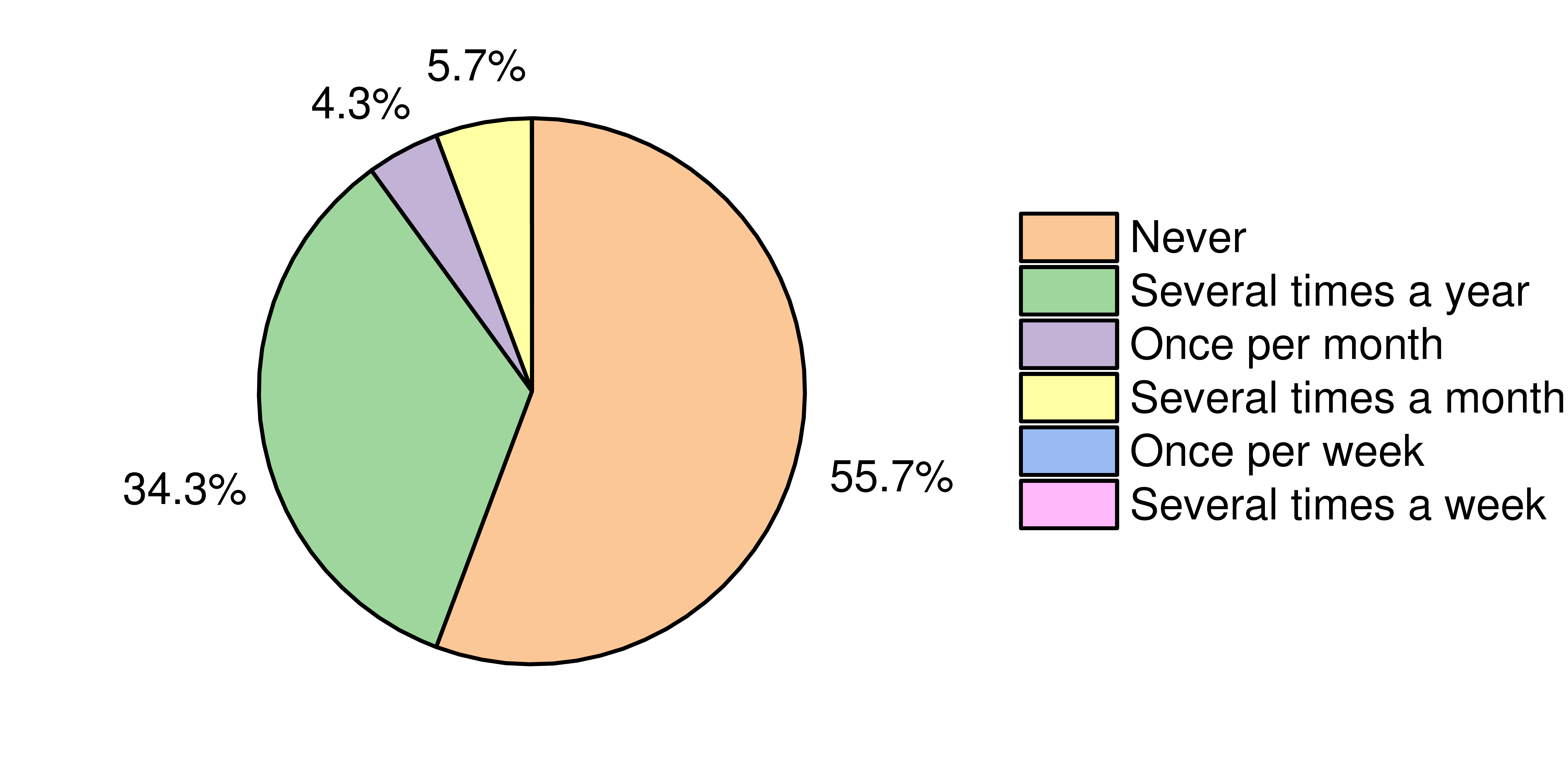}
\caption{Answers to the question \textit{``How often do you use the Extract Method refactoring feature in the IDE?''}}
\label{fig:extractmethodusage_refactoring}
\end{figure}

We started the survey by asking participants how often they use the \textit{Extract Method} refactoring feature in the IDE. Figure~\ref{fig:extractmethodusage_refactoring} shows the breakdown of the answers. It can be seen that \textit{Extract Method} refactoring is not very frequently used in the IDE as 55.7\% of developers indicated that they never used the \textit{Extract Method} refactoring feature, while 34.3\% said they used it several times a year. Only 4.3\% of the users said that they extracted methods approximately once per month, and just 5.7\% perform \textit{Extract Method} several times a month. It appears that not many developers have used the IDE to apply the \textit{Extract Method} refactoring. This observation corroborates the finding of a recent survey on refactoring \cite{golubev2021one} that shows significantly fewer developers using the IDE feature for the \textit{Extract} refactorings, especially when compared to the \textit{Rename} feature. The fact that \textit{Extract} is not as intuitive as \textit{Rename} represents a significant challenge to the ongoing research and studies that develop recommendation algorithms to extract code at any level (method, class, package, etc.). This is one of the main motivations behind the design of our solution. We believe that our tool can further support developers to practice this type of refactoring, or at least raise their awareness of its existence regardless.

\begin{figure}
\centering 

\includegraphics[width=1.0\columnwidth]{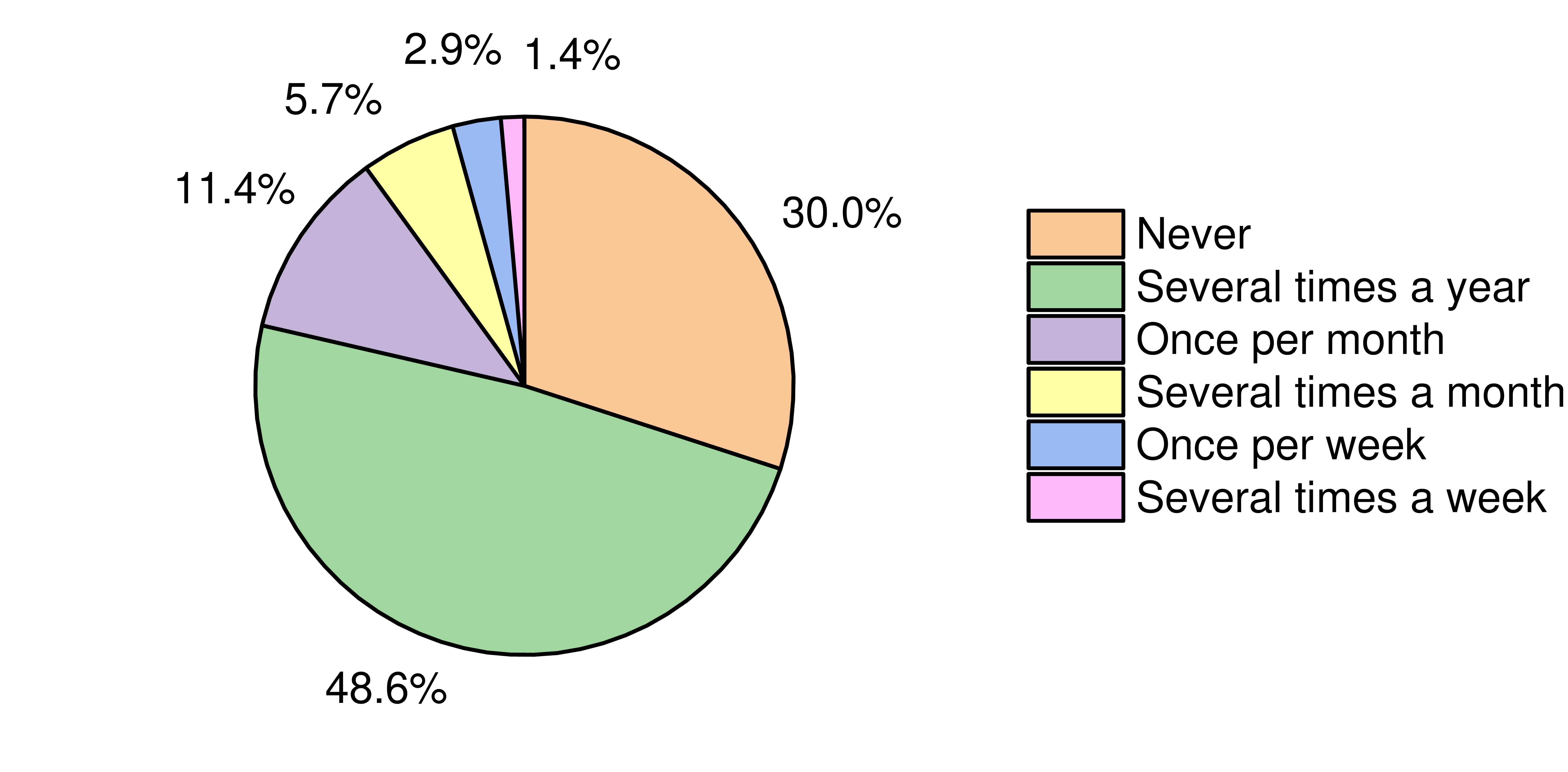}
\caption{Answers to the question \textit{``How often have you refactored duplicate code / code clones?''} 
}
\label{fig:codeclone_refactoring}
\end{figure}

Concerning the frequency of refactoring specifically duplicated code, Figure~\ref{fig:codeclone_refactoring} depicts that almost half of the developers answered that they refactor duplicated code several times a year. 30\% of the respondents said that they never refactor duplicated code. 11.4\% of the respondents said that they refactor duplicated code once per month. We conjecture that despite the existence of automated code clone detectors (\eg CCFinder~\cite{kamiya2002ccfinder}), these tools might lack integration, as developers acknowledge the existence of code clones but they do not have a preference on how to refactor them using automated tools. Besides, several studies have also shown that not all duplicates are harmful to the code~\cite{mens2021good}.

\begin{table}[t]
\centering
\caption{The results of tracking features in collected logs.}
\label{Table:tracking feature}
\begin{tabular}{@{}ll@{}}
\toprule
\textbf{Event} & \textbf{Count} \\ 
\midrule
 notificationCount & 63 \\
 extractMethodAppliedCount & 59 \\
 extractMethodCanceledCount & 0 \\
 copyCount & 350 \\
 pasteCount & 379 \\
 \bottomrule
\end{tabular}
\end{table}

Table \ref{Table:tracking feature} reveals the sum of metric numbers, extracted from the log files shared by the participants. Our first observation is that only 17\% of the pasted code was evaluated as worth refactoring by the model. This can be due to the CNN being \textit{picky} by nature, which explains that 93\% of its recommendations were actually accepted by the participants. Since the number of cancelled refactorings is zero, we conjecture that the 7\% of cases where the model recommends a refactoring that was not applied, can be due to developers simply ignoring the notification shown in their screen. This is another advantage of the pop-up notification --- it has a minimal disturbance on the developers when they are busy.

\begin{table*}[htbp]
  \centering
	 \caption{Developer's insight about the usefulness,  usability, and functionality of the tool.}
  \vspace{0.5cm}
	 \label{Table:example}
\begin{sideways}
\begin{adjustbox}{width=1.05\textwidth,center}
\begin{tabular}{llLllll}\hline
\toprule
\bfseries Category & \bfseries Sub-category & \bfseries Example (Excerpts from a related survey response) \\
\midrule
\multirow{16}{*} {\textbf{Usefulness}} & \cellcolor{gray!30} {Effort}  & \cellcolor{gray!30} \say{\textit{I feel it should be useful as this \textbf{saves time and effort} put but the developer who is developing a software.}} \\ 
&  Quality & \say{\textit{A plugin that automatically extracts methods from duplicated code sounds very useful for the sake of \textbf{reducing complexity and improving readability of code}. In my own coding experience, I've definitely had times where I recognized that I was duplicating code, but the means by which the method should be extracted wasn't immediately obvious.}} \\ 
& \cellcolor{gray!30} {Automation} & \cellcolor{gray!30} \say{\textit{For small code projects, I don't think it would be as useful, but for large code projects it probably would be. A suggestion, to make the tool even more useful, would be to allow for static analysis of large code bases (where the tool would be most useful) that have already been coded, then recommending extract method refactorings. Companies with large messy code bases would love to have a tool like that to clean up all their code duplicates \textbf{without having to meticulously go through it manually}.}} \\ 
& Awareness &  \say{\textit{As a developer, when working with large files and thousands of code lines we tend to repeat some of the functions already present in the code which results to duplicity and less efficient. This tool is very helpful and \textbf{makes the developer aware of their approach} and not to make the mistakes of repeating the code again and again.}} \\ 
\hline
\multirow{10}{*} {\textbf{Usability}} & \cellcolor{gray!30} Notification & \cellcolor{gray!30} \say{\textit{I guess adding a notification was great touch to the project. but I would rather \textbf{have an option to enable or disable the notification}. Plus i would also like it to highlight the duplications in the code itself rather then a notification.}} \\ 
&  Delay & \say{\textit{I would \textbf{remove the delay} and just put a UI notification in bottom that refactoring is available. If the user tries to edit the code, then the notification disappears.}} \\ 
& \cellcolor{gray!30} Dialog box & \cellcolor{gray!30} \say{\textit{I would \textbf{remove the dialog} as it can interrupt a users workflow. I would set it as an alert in the UI such that it doesn't interrupt a user but rather notifies them and gives the ability to make the fix or easily continue developing with the duplicate code.}} \\ 
&  Preview &  \say{\textit{A change in the operation of the plugin would be \textbf{showcase a preview of the code itself instead of just the signature preview}. Seeing the code change as you name the new function would be a nice change.}} \\ 
\hline
\multirow{5}{*} {\textbf{Functionality}} & \cellcolor{gray!30} Recommendation & \cellcolor{gray!30} \say{\textit{Perhaps \textbf{give a default name to the function} instead of asking user to enter manually.}} \\
&  Refactoring &  \say{\textit{Instead of making a pop-up appear whenever duplicated code is detected, I would prefer to \textbf{have an option to scan for instances of duplicated code} and have it automatically refactored after that.}} \\ 
\bottomrule
\end{tabular}
\end{adjustbox}
\end{sideways}
\end{table*}

In Table~\ref{Table:example}, we report the main thoughts, comments, and suggestions about the overall impression of the usefulness, usability, and functionality of the proposed idea, in accordance with the conducted labeling. Table~ \ref{Table:example} also presents samples of the participants' comments to illustrate their impressions.

\smallskip
\textbf{Usefulness.} Generally, the respondents found the tool to be useful in regard to four main aspects: effort, quality, automation, and awareness. The majority of the participants commented that the proposed idea saves time and effort for developers who would not have to examine and refactor duplicates manually. Other participants communicated that reducing redundant code assists in increasing its readability and efficiency while reducing its complexity, which helps improve overall code quality. A moderate subset of developers revealed that the tool's ability to identify duplicates within a file and reduce them to a single method allows users to only correct errors in a specific location pro-actively and automatically. Further, some developers commented that the tool aids developers in identifying their redundancy when updating a source file that they are not familiar with. Another noteworthy point mentioned was that the tool helps less experienced and novice coders in writing well-structured code.

\smallskip
\textbf{Usability.}  Based on the feedback provided by the respondents to the survey, the key areas in usability related to the notification, the delay, the dialog box, and the preview. The notification and the delay aspects of the tool were the primary areas identified for change or improvement. The main suggestions about the pop-up notification were driven by developers' personal preferences of how notifications should be shown. For example, some respondents viewed the current pop-up notification (Figure~\ref{fig:example}) as something that could be distracting, and suggested other options. These options included highlighting the duplicate code, icon flashing at the bottom taskbar, or just using a warning message. Some responses stated that the users of the tool should have more control to set the delay time to wait before triggering the suggestion. Other responses also stated that triggering the tool after a save operation would be less disruptive. While many of these ideas can be added to tailor the tool to developers' preferences, we find the suggestion of highlighting duplicate code to be the most practical and we plan to implement it in the future.

\smallskip
\textbf{Functionality.} From the participants' feedback, we have also extracted suggestions to improve the tool's functional features. One proposed feature was to recommend the proper method name for the extracted code, based on its functionality. Moreover, since the participants found the tool to be useful, they suggested to support additional refactorings to remove duplicates at the class level (\ie \textit{Extract Class}). Participants also suggested scanning all project files for duplicates (instead of just the current one, as it is now implemented) and then interactively suggest their refactoring one by one. We found this suggestion to be particularly interesting, since it does not match the rationale of our solution. Implementing this suggestion would require training with positive samples that were exclusively executed to remove duplicates. It would also require taking into account the imbalanced nature of the codebase by creating a significantly higher number of negative samples.

\begin{figure}[t]
\centering

\includegraphics[width=.8\columnwidth]{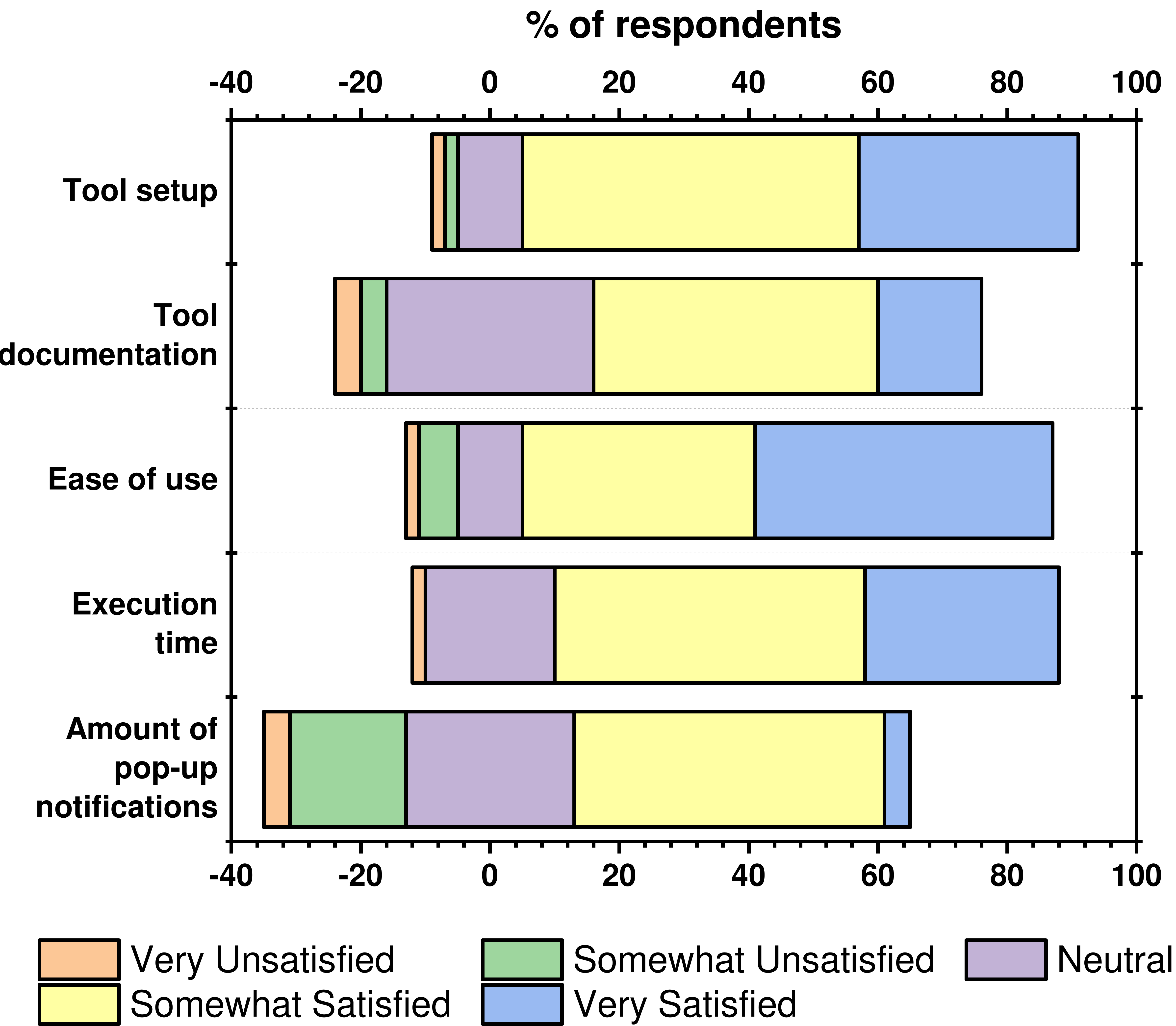}
\caption{Participants' satisfaction with various  aspects of the \toolname tool.}
\label{fig:likertscale}
\end{figure}

50 out of 72 participants confirmed that they executed the plugin. We asked them about their satisfaction with various aspects of the tool. Figure~\ref{fig:likertscale} presents an overview of their answers. With respect to the tool setup, most of the respondents (43 participants) reported that they are satisfied with the tool. Regarding the tool documentation, the majority of the respondents agreed that the documentation is useful; only 4 participants were unsatisfied. For the ease of use aspect, a larger group (41 participants) was satisfied. Several participants found that the tool is not easy to use, so we will work on improving its usability.  Concerning the execution time, 39 participants were happy with it. For the amount of pop-up notifications, the majority of respondents (26 participants) agreed that the amount of pop-up notifications is acceptable, although there were a few participants that remained neutral. There were also a few participants who were not happy with the amount of pop-up notifications, and we are planning on improving this aspect of the tool in the future.

\begin{tcolorbox}
\textit{\textbf{Summary:} Overall, the participants were satisfied with the plugin and rated its various aspects highly. The aspects of the tool that can be improved are the UI of highlighting and more settings for users' preferences.} 
\end{tcolorbox}

\section{Threats to Validity}
\label{Section:Threats}
In this section, we identify potential threats to the validity of our approach and our experiments as discussed in the work of Ampatzoglou~\etal~\cite{ampatzoglou2019identifying}.

\textbf{Internal Validity.} Our analysis is mainly threatened by the accuracy of the refactoring mining tool because the tool may miss the detection of some refactorings. However, previous studies \cite{tsantalis2018accurate,tsantalis2020refactoringminer} report that RefactoringMiner has high precision and recall scores (99.8\% and 95.8\%, respectively) compared to other state-of-the-art refactoring detection tools, which gives us confidence in using this tool. Also, since the developers have the option to just "skip" the notification, there were no cases when developers started the process and then rejected it. It is also possible that developers undid the refactoring after conducting it. 

\textbf{Construct Validity.} Collecting positive examples for our model requires not only finding the refactoring itself but navigating to the previous commit to see the context, from where the method was extracted. In some cases, the detection of the previous commit might not be straightforward because there are several branches in the repository. We found several such cases when manually checking the collected data. As for the collection of negative examples, it was done using the ranking algorithm inspired by the one of Haas and Hummel \cite{haas2016deriving}, so our work inherits any limitation associated with that algorithm. Concerning the completeness and correctness of our interpretation of the open-ended responses within the survey, we did not extensively discuss all responses because some of them are open to various interpretations, and we need further follow-up surveys or interviews to clarify them.

\textbf{External Validity.} Our analysis was performed on 13 mature open-source Java projects belonging to the Apache ecosystem that are varied in size, contributors, number of commits and refactorings. However, we cannot claim the generality of our observations to projects written in other programming languages or belonging to other ecosystems. Further investigation of even more projects is needed to mitigate this threat. Regarding the study participants, the majority of our participants involved students, with some of them having industrial experience. To avoid bias in the experiment, we make providing feedback anonymous and not mandatory, to increase the magnitude of tool usage experience. Although feedback was optional, 95\% of students have completed it after removing arbitrary submissions. As future work, we plan to perform another round of external validation with professional software engineers in industry to hear their perception.

\section{Conclusion}
\label{Section:Conclusion}

Recommending \textit{Extract Method} refactoring opportunities is of paramount importance to the research community and industry. Although a plethora of studies have utilized a variety of approaches to identify \textit{Extract Method} refactoring, recommending this refactoring type without 
interfering with developers' workflow remains largely unexplored. In this study, we proposed \toolname as an IntelliJ IDEA plugin, and experimented with machine learning models in order to increase the adoption and usage of the \textit{Extract Method} refactoring while maintaining the workflow of a developer. Our results reveal that machine learning models are able to recommend \textit{Extract Method} refactoring opportunities as soon as code duplicates are introduced in the IDE, and the participants were satisfied with the \toolname tool.

In particular, the proposed CNN demonstrated an F-measure of 0.82 and outperformed other machine learning models. In the survey, we discovered that a majority of developers carry out \textit{Extract Method} refactorings very rarely or never at all, so the proposed pro-active pipeline for their recommendation can also fulfill the educational needs. Overall, the participants rated various aspects of the plugin highly, while also providing valuable ideas for future development. In particular, we would like to implement the highlighting of refactorable code duplicates in the editor and give the users more control over various aspects of the plugin for customization.

{\footnotesize\bibliography{mybibfile}}

\end{document}